\documentclass[epj]{webofc}
\usepackage[varg]{txfonts}   
\wocname{EPJ Web of Conferences}
\woctitle{CONF12}

\usepackage{slashed}
\usepackage{amsmath,graphicx}
\usepackage[colorlinks=true,linktocpage=true,linkcolor=blue,citecolor=blue]{hyperref}
\usepackage{float}
\usepackage{nicefrac}
\usepackage[normalem]{ulem}
\usepackage{amsmath}
\usepackage{subcaption}

\def\p{{\boldsymbol p}}
\def\x{{\boldsymbol x}}
\def\v{{\boldsymbol v}}
\def\xT{{\boldsymbol x}_T} 
\newcommand{\RAA}{R_\text{AA}}
\newcommand{\Np}{N_\text{part}}
\newcommand{\pT}{p_T}
\newcommand{\pL}{p_L}
\newcommand{\feq}{f_\text{eq}}
\newcommand{\fiso}{f_\text{iso}}
\newcommand{\fa}{f_\text{aniso}}
\newcommand{\teq}{\tau_\text{eq}}
\newcommand{\Nd}{N_\text{dof}}
\newcommand{\aniso}{\text{aniso}}
\newcommand{\iso}{\text{iso}}
\newcommand{\eq}{\text{eq}}
\newcommand{\Eb}{E_\text{bind}}
\newcommand{\gd}{\gamma_\text{dis}}

\newcommand{\ramuno}{i} 

\begin{document}
\selectlanguage{english}
\title{
Bottomonia suppression in an anisotropic quark-gluon plasma 
}
%
%

\author{Radoslaw Ryblewski\inst{1}\fnsep
\thanks{\email{Radoslaw.Ryblewski@ifj.edu.pl}}
}

\institute{Institute of Nuclear Physics, Polish Academy of Sciences, PL-31342 Krak\'ow, Poland
}

\abstract{A brief review of recent studies on suppression of bottomonia in an anisotropic quark-gluon plasma created in heavy-ion collisions at the LHC is presented. A reasonable agreement between the model predictions for the inclusive $R_{\rm AA}$ suppression factor and the preliminary CMS experimental data is found. The values of the shear viscosity to the entropy density ratio extracted from the comparison with the data lie between one and two times the gauge/gravity duality lower bound. These values agree very well with the fluid dynamical fits to the light hadron correlation data and confirm that the quark-gluon plasma is a nearly-perfect fluid.

}
\maketitle
%
\section{Introduction}
\label{sec-1}
%
\par One of the main objectives of the ongoing ultra-relativistic heavy-ion collisions (URHIC) studies at the Large Hadron Collider (LHC) in CERN is to produce and extract the properties of a new state of hot nuclear matter called quark-gluon plasma (QGP) \cite{Florkowski:2010zz,Gale:2013da,Jaiswal:2016hex}. An enormous amount of fluctuation and correlation data collected in URHIC suggests that the bulk of the created matter is a strongly-coupled  system, which, to a great extent, behaves as a nearly-perfect relativistic fluid with the temperature-averaged shear viscosity to entropy density ratio, $\eta/s$, in the range $1/(4\pi) - 3/(4\pi)$ \cite{Gale:2013da,Jaiswal:2016hex}. The space-time evolution of the latter may be quite precisely described within the recent formulations of second-order relativistic viscous hydrodynamics \cite{Denicol:2012cn,Jaiswal:2013npa}. Lately, it was also shown that the description of the very early evolution times requires careful treatment of potentially large local momentum anisotropies in the system. For that purpose one should rather use an alternative fluid dynamics approach called anisotropic hydrodynamics in its leading \cite{Florkowski:2010cf,Martinez:2010sc,Martinez:2010sd,Ryblewski:2010bs,
Ryblewski:2011aq,Ryblewski:2012rr,Martinez:2012tu,Tinti:2013vba,Nopoush:2014pfa,Tinti:2014yya,Nopoush:2015yga,Tinti:2015xwa,Alqahtani:2015qja,Nopoush:2016qas} or next-to-leading \cite{Bazow:2013ifa,Molnar:2016vvu} order formulation.
\par The first-principles calculations of the quantum chromodynamics (QCD) equation of state within Hard Thermal Loop (HTL) approach \cite{Haque:2014rua,Mogliacci:2013mca} suggest that the high-temperature QGP may be well described  as a decoupled system of partonic quasiparticles with the pseudocritical transition temperature \mbox{$T_c\approx 165$ MeV}. One of the signatures of QGP creation is the ``melting'' of heavy hadronic states due to the Debye screening phenomenon, commonly measured with respect to the production in p-p collisions with the $\RAA$ suppression factor. The fits of the relativistic viscous fluid models to the light hadron production data suggest that at the top LHC energies the  system created in central collisions reaches a peak temperature on the order of $T_0\approx 600$ MeV \cite{Gale:2013da}. As a result, the light hadronic states, which mostly disassociate already around $T_c$, provide a limited source of information on the  properties of the hottest part of the medium. At the same time it was shown that heavy quarkonium states, such as bottomonia, may survive up to temperatures on the order of $\approx 4 T_c$ \cite{Mocsy:2013syh,Andronic:2015wma}, which makes them a potential probe of the early stages/very center of the created fireball. Moreover, due to their sequential disassociation pattern, they provide a possibility to distinguish between different stages of the QGP evolution \cite{Karsch:2005nk}. 
\par In this proceedings contribution we briefly review our main results on the thermal suppression of $\Upsilon(1s)$ and $\Upsilon(2s)$ bottomonium states in an anisotropic QGP created in $2.76$ TeV Pb-Pb collisions at the LHC \cite{Krouppa:2015yoa}. For this purpose, we use an updated potential-based non-relativistic QCD (pNRQCD) model developed in Refs.~\cite{Strickland:2011mw,Strickland:2011aa} coupled to the anisotropic hydrodynamics model for the background evolution constructed in Ref.~\cite{Ryblewski:2015hea}. The updates to the model include: (a) realistic (3+1)-dimensional QGP evolution within anisotropic hydrodynamics, (b) update of the mixing fractions of different bottomonia states based on the recent ATLAS, CMS, and LHCb measurements, and (c) improved centrality averaging procedure. Herein we show selected results on the inclusive $\RAA$ suppression factor of the $\Upsilon$ states as a function of the number of participants, $\Np$, and transverse momentum, $\pT$, and compare them to the preliminary CMS results \cite{CMS:2011ora}. We observe that the employed model provides a satisfactory description of the experimental data. Moreover, the restriction on the values of the shear viscosity to the range $1/(4\pi)<\eta/s<2/(4\pi)$ extracted from the comparison to the data agree quite well with the results of the fluid dynamical fits to the light hadron correlation data. The latter confirms creation of an almost perfect QGP at LHC energies.
\section{Spheroidally momentum-anisotropic QGP}
\label{sec-2}
%
\par Most of the microscopic models of the QGP early-time dynamics suggest that, although QGP is initially highly-anisotropic in the momentum space \cite{Ryblewski:2013jsa,Strickland:2014pga}, its evolution closely follows the dissipative fluid dynamics equations \cite{Heller:2011ju,Jankowski:2014lna}. One can show that, due to the specific topology of the URHIC, in particular rapid expansion along the beam ($z$) direction and relatively slow expansion in the direction transverse to it, the dominant dissipative corrections to the isotropic single-particle phase-space distribution function,
\begin{equation}
\fiso(p^\mu,x^\mu)= \fiso\left(|\p|,T(x^\mu)\right),  
\label{feq}
\end{equation}
follow from the anisotropy between the transverse, $p_T^2\equiv p_x^2+p_y^2 $, and longitudinal, $\pL^2\equiv p_z^2$, direction in momentum space and result in a significant difference between the transverse and longitudinal pressure, $P_T \gg P_L$. Here we take  $x^\mu=(t,\x)$ and $p^\mu=(E,\p)$ where $E=\sqrt{m^2+\p^2}$ is the particle on-shell energy. The most straightforward way to include these corrections is to use the following spheroidal Romatschke-Strickland (RS) form  of particle distribution function \cite{Romatschke:2003ms}
\begin{equation}
\fa(p^\mu,x^\mu)=  \fa\left(\p,\xi (x^\mu),\Lambda (x^\mu)\right) = \fiso\left(\sqrt{\pT^2 + \left(1+\xi(x^\mu)\right)\pL^2},\Lambda(x^\mu)\right),
\label{faniso}
\end{equation}
where $-1 \leq \xi(x^\mu) < \infty$ is the momentum-space anisotropy parameter, $\Lambda(x^\mu)$ is the transverse temperature and $\fiso$ is an arbitrary isotropic distribution function. Hereafter it is assumed that the underlying parton distribution function, $\fa(p^\mu,x^\mu)$, is the same for the QGP background evolution as well as for the quarkonium binding calculations.
%
\section{Fluid dynamics of the anisotropic background}
\label{sec-3}
%
\subsection{Evolution equations}
\label{sec-3-1}
%
\par With the knowledge of the single-particle distribution function one can derive equations of motion for the soft modes of the system using the standard relativistic kinetic theory formalism. In particular, the RS form (\ref{faniso}) of the distribution function leads to the fluid dynamical equations of the so called leading-order anisotropic hydrodynamics \cite{Florkowski:2010cf,Martinez:2010sc,Martinez:2010sd,Ryblewski:2010bs}. In the formulation employed in the present study they are obtained by taking the lowest-$n$ momentum moments \cite{Martinez:2010sc,Martinez:2010sd,Martinez:2012tu}, 
\begin{equation}
{\cal \hat{I}} ^{\mu_1\cdots\mu_n}\equiv  \int \!dP\, p^{\mu_1}p^{\mu_2}\cdots p^{\mu_n}, \qquad \qquad 
\int dP \equiv \frac{\Nd}{(2\pi)^3}\int \frac{d^3\!\p}{E},
\end{equation}
of the Boltzmann equation,
\begin{equation}
p^\mu \partial_\mu f=-{\cal C}[f],
\end{equation}
with the collisional kernel treated in the relaxation-time approximation, 
\begin{equation}
{\cal C}[f]=p_\mu u^\mu(f-\feq(|\p|,T(x^\nu)))/\teq.
\end{equation}
In the above definitions $\Nd$ denotes the number of degrees of freedom, $\teq$ is the relaxation time, and $\feq$ is the equilibrium distribution which we take in the Boltzmann form $\feq(|\p|,T) = \exp(-|\p|/T)$. Within the spheroidal ansatz (\ref{faniso}), $f=\fa$, it is sufficient to restrict to the zeroth and the first moments which leads to the following set of dynamical equations, 
\begin{equation}
\partial_\mu {\cal I}_\aniso^\mu = u_\mu({\cal I}^\mu_\eq-{\cal I}_\aniso^\mu)/\teq,
\label{ce}
\end{equation}
\begin{equation} 
\partial_\mu {\cal I}_\aniso^{\mu\nu}=u_\mu ({\cal I}^{\mu\nu}_\eq-{\cal I}_\aniso^{\mu\nu})/\teq,
\label{emc}
\end{equation}
where  we defined ${\cal I}_\text{aniso, eq}^{\mu_1\cdots\mu_n}\equiv {\cal \hat{I}} ^{\mu_1\cdots\mu_n} f_\text{aniso, eq}$. In Eq.~(\ref{emc}) one may also make use of the energy-momentum conservation, which leads to the  Landau matching condition, $u_\mu {\cal I}^{\mu\nu}_{\rm eq}=u_\mu {\cal I}_\aniso^{\mu\nu}$, where $u^\mu=\gamma(1,{\bf v})$ is the fluid four-velocity defined in the Landau frame with $u^\mu u_\mu =1$.  The forms of particle four-current, $N_\text{aniso, eq}^\mu\equiv {\cal I}_\text{aniso, eq}^{\mu}$, and the energy-momentum tensor, $T_\text{aniso, eq}^{\mu\nu} \equiv {\cal I}_\text{aniso, eq}^{\mu\nu}$, may be found through their tensor decomposition in a suitable orthonormal four-vector basis $\{I\}_\text{LRF}$, where $I \in \{u, X, Y, Z\}$ \cite{Martinez:2012tu}. In this way one obtains \cite{Florkowski:2010cf,Martinez:2010sc,Martinez:2010sd,Ryblewski:2010bs,
Ryblewski:2011aq,Ryblewski:2012rr,Martinez:2012tu,Tinti:2013vba,Nopoush:2014pfa,Tinti:2014yya,Nopoush:2015yga,Tinti:2015xwa,Alqahtani:2015qja,Nopoush:2016qas} 
\begin{equation} 
N^\mu = n_\text{aniso, eq} u^\mu,
\label{N}
\end{equation}
\begin{equation} 
T_\text{aniso, eq}^{\mu\nu} = \varepsilon^\text{aniso, eq} u^\mu u^\nu + P_T^\text{aniso, eq} (X^\mu X^\nu+Y^\mu Y^\nu)+ P_L^\text{aniso, eq} Z^\mu Z^\nu,
\label{TMN}
\end{equation}
see also \cite{Florkowski:2008ag,Florkowski:2009sw}. Projecting (\ref{N}) and (\ref{TMN}) on the basis four-vectors and performing momentum integrals one obtains the explicit form of the thermodynamic variables entering Eqs.~(\ref{N}) and (\ref{TMN})
\begin{equation} 
n^\aniso(\Lambda,\xi) 
=  n^\iso(\Lambda)/ \sqrt{1+\xi},
\end{equation}
\begin{equation} 
\varepsilon^\aniso(\Lambda,\xi) 
= {\cal R}(\xi) \varepsilon^\iso(\Lambda),
\end{equation}
\begin{equation}
P_T^\aniso(\Lambda,\xi)={\cal R}_{T}(\xi)P_T^\iso(\Lambda),  
\end{equation}
\begin{equation}
P_L^\aniso(\Lambda,\xi) = {\cal R}_{L}(\xi)P_L^\iso(\Lambda),
\end{equation}
where $P_T^\iso = P_L^\iso = P^\iso$, and ${\cal R}$ are known analytic functions \cite{Martinez:2010sc}. Transforming to the laboratory frame and expanding equations (\ref{ce}) and (\ref{emc}) one obtains the explicit equations of motion for $\Lambda (\tau,\xT, \varsigma)$, $\xi(\tau,\xT, \varsigma)$ and $\v(\tau,\xT, \varsigma)$ \cite{Ryblewski:2015hea}, where $\tau=\sqrt{t^2 -z^2}$ is the longitudinal proper-time and $\varsigma = {\rm tanh}^{-1}(z/t)$ is the space-time rapidity. Following the usual methodology we take $\fiso=\feq$ and restrict ourselves to the massless case, which gives the well known relations for the conformal system, $\varepsilon^\text{iso}(T) =3P^\text{iso}(T) = 3 T n^\text{iso}(T) \sim T^4$.
%
\subsection{Setup}
\label{sec-3-2}
%
\par For the numerical simulations of the hydrodynamic background evolution we choose the initial proper time $\tau_0 =0.3$ fm. The values of the initial central temperature in central collisions $T_0 \in \{552, 546, 544\}$ MeV are determined for each value of $4 \pi \eta/s \in \{1, 2, 3\}$, respectively, in order to reproduce the final charged-particle multiplicity measured in the experiment. The initial profile for the transverse temperature $\Lambda (\tau=\tau_0,\xT, \varsigma)$ in the transverse plane follows from the initial energy density profile which is set by the mixed optical Glauber model with mixing factor $\kappa_{\rm binary} = 0.145$. The inelastic cross-section is taken to be $\sigma_{NN} = 62$ mb. In the spatial rapidity direction, we use the following phenomenological distribution consistent with the limited fragmentation picture at large rapidity \cite{Bozek:2012qs}
\begin{equation}
f(\varsigma) \equiv \exp \!\left[ - \frac{(\varsigma - \Delta \varsigma)^2}{2 \sigma_\varsigma^2} \Theta (|\varsigma| - \Delta \varsigma) \right].
\label{longprof}
\end{equation}
In Eq.~(\ref{longprof}) the parameters $\Delta\varsigma = 2.5$ and $\sigma_{\varsigma} = 1.4$ are fitted to reproduce the pseudorapidity distribution of light charged hadrons measured in the experiment and $\Theta$ is the Heaviside step function.
%
\section{Quarkonium potential in anisotropic QGP}
\label{sec-4}
%
\par Following findings of the HTL resummed perturbation theory calculations \cite{Laine:2006ns} it is assumed that the static heavy quarkonium potential is complex valued, $V=\Re[V]+ \ramuno\Im[V] $. For the real part of the potential one uses the results obtained for a finite-temperature spheroidally anisotropic QGP \cite{Dumitru:2009ni} described by the distribution function (\ref{faniso}). It is based on the internal energy of the states calculated from the Karsch-Mehr-Satz form \cite{Karsch:1987pv} of the free energy \cite{Strickland:2011aa}, 
\begin{equation}
{\Re}[V] = -\frac{a}{r}(1+\mu) e^{-\mu r} + \frac{2 \sigma}{\mu}(1-e^{-\mu r}) - \sigma r e^{-\mu r} -\frac{0.8 \sigma}{m_b^2 r},
\label{ReV}
\end{equation} 
where $a\equiv \alpha_s C_F=0.385$ is treated as a parameter which is fitted to the lattice QCD data to reproduce screened Coulomb part of quarkonium potential \cite{Petreczky:2010yn}, $\sigma = 0.223 {\rm\, GeV^2}$ is the string tension also obtained from lattice QCD calculations \cite{Petreczky:2010yn}, and $m_b=4.7 {\rm\, GeV}$ is the constituent mass of the bottom quark. The anisotropic screening mass $\mu = {\cal G}(\xi, \theta) m_D$  is expressed through the isotropic Debye mass $m_D = 1.4 \sqrt{1+N_f/6}g_s \Lambda$ and the function $\cal G$ which depends on the  anisotropy $\xi$ and the $\theta$ angle between the beam line direction and the line connecting $Q{\bar Q}$ pair  \cite{Strickland:2011aa}. The factor $1.4$ is included to account for higher-order corrections determined from lattice simulations \cite{Kaczmarek:2004gv}, $N_f =2$ is the number of quark flavors contributing to the medium, and $g_s= \sqrt{4 \pi \alpha_s}$, where we take three-loop-running strong coupling giving $\alpha_s(5 {\rm\, GeV}) = 0.2034$.

The imaginary part of the potential is obtained using leading-order perturbative calculation performed in the small anisotropy limit \cite{Dumitru:2009fy},
\begin{equation}
{\Im}[V] =  -\alpha_s C_F \Lambda \left\lbrace\vphantom{frac{1}{2}}\phi\left(\mu r\right)-\xi \left[ \psi_1 \left(\mu r,\theta\right)+ \psi_2 \left(\mu r,\theta\right)\right]\right\rbrace,
\label{ImV}
\end{equation}
where $C_F=4/3$ is the QCD color factor, and $\phi$ and $\psi$ are special functions expressible in terms of the Meijer G-function.
%
\section{Quarkonium local decay rate}
\label{sec-5}
%
\par
\begin{figure}[t!]
\centering
\includegraphics[width=7cm,clip]{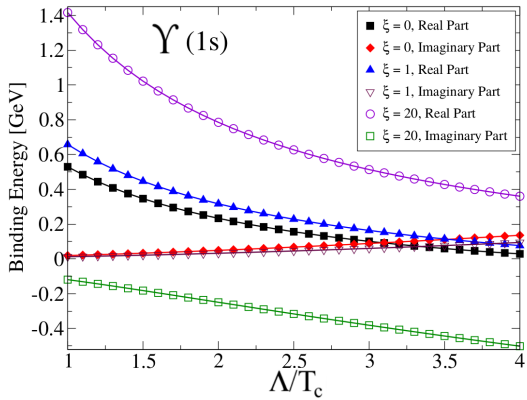} 
\caption{(Color online) Real and imaginary parts of the $\Upsilon(1s)$ binding energy as a function of $\Lambda/T_c$ ratio for various values of $\xi$, see \cite{Strickland:2011aa} for details.}
\label{fig-1}       
\end{figure}

 Using the form of the quarkonium potential defined in Eqs.~(\ref{ReV})-(\ref{ImV}) we solve the three-dimensional Schr\"odinger equation for the complex-valued binding energies of the states, $\Eb$, as functions of bulk variables $\xi$ and $\Lambda$ \cite{Margotta:2011ta}. The exemplar results for binding energy of $\Upsilon(1s)$ state are shown in the Fig.~\ref{fig-1}. One may observe that in the anisotropic system the disassociation point, defined as $\Lambda$ at which ${\Im}[\Eb]={\Re}[\Eb]$, is shifted towards higher $T$, which means that the anisotropy $\xi>0$ causes the states to melt at higher transverse temperatures.
%
%
\par While the real part of the binding energy defines if the state is bound ($\Eb \le 0$) or not  ($\Eb >0$), the imaginary part gives the information about the local decay rate (width) $\Gamma$ of the state. Computing the quantum mechanical occupation number one obtains the relation
\begin{equation}
\Gamma(\Lambda, \xi) = 
\left\{
\begin{array}{ll}
2 \Im[\Eb]  & \;\;\;\;\; \Re[\Eb] >0 \\\nonumber
\gd = 10\,{\rm GeV}  & \;\;\;\;\; \Re[\Eb] \le 0 \nonumber
\end{array}
\right. ,
\label{width}
\end{equation}
where $\gd = 10\,{\rm GeV}$ is phenomenological parameter which sets the decay rate of the unbound states.
\par Knowing $\Eb(\xi,\Lambda)$ from the solution of the Schr\"odinger equation and the space-time dependence $\Lambda (\tau,\xT, \varsigma)$ and $\xi(\tau,\xT, \varsigma)$ from the hydrodynamic evolution of the QGP (see Sec.~\ref{sec-3}), one may construct  $\Eb(\tau,\xT, \varsigma)$.  
%
\section{$\text{R}_{\text{AA}}$ suppresion factor}
\label{sec-5}
%
\par The $\RAA$ suppression factor is obtained by the integration of the local decay rate,
\begin{equation}
\zeta (\pT,\xT,\varsigma) \equiv \Theta(\tau_\text{f}-\tau_\text{form}) \int_{\text{max}(\tau_\text{form},\tau_\text{0})}^{\tau_\text{f}}\hspace{-1cm} d\tau\,\Gamma(\tau,\xT,\varsigma),
\label{zetaraa}
\end{equation}
and subsequent exponentiation of the result, $\RAA(\pT,\xT,\varsigma) = \exp\left(-\zeta(\pT,\xT,\varsigma)\right)$.
The formation time of the state is $\tau_\text{form}(\pT) = \tau_\text{form}^0 \gamma  =  \tau_\text{form}^0 E_T /M$ with $M$ being the mass of the state and $E_T=\sqrt{\pT^2+M^2}$ its transverse energy. The rest-frame formation times $\tau_{\rm form}^0$ are assumed to be inversely proportional to the vacuum binding energy \cite{Karsch:1987uk}, which gives $\tau_{\rm form}^0 = 0.2, 0.4, 0.6, 0.4, 0.6$ fm for $\Upsilon(1s)$, $\Upsilon(2s)$, $\Upsilon(3s)$, $\chi_{b1}$, $\chi_{b2}$ states, respectively. The final time $\tau_\text{f}$ is defined through the condition $T(\tau_\text{f},\xT,\varsigma) \le T_c$, where $T=\Lambda{\cal R}^{1/4}(\xi)$. For the space-time dynamics of the quarkonia themselves we use a simplistic assumption that, once generated, they follow the Bjorken flow solution, which means that hereafter we may put $\varsigma = y$. Integration over the spatial coordinates 
\begin{equation}
\RAA(\pT,y) = \frac{\int d^2\!\xT n(\xT,y) \RAA (\pT,\xT,y) }{\int  d^2\!\xT n(\xT,y)}, 
\label{zetaraa}
\end{equation}
takes into account the fact that the generation of the quarkonia is assumed to be proportional to the local number
density of plasma partons $n(\xT,\varsigma=y) = n^\aniso(\Lambda(\tau_\text{0},\xT,\varsigma),\xi(\tau_\text{0},\xT,\varsigma))$.
\par Before performing comparison with the data one has to perform certain averages of the $\RAA$ taking into account proper momentum cuts according to the experimental ones. 
For the transverse-momentum average we use the $E_T^{-4}$ measured by CDF \cite{Acosta:2001gv} at high $\pT$
\begin{equation}
\RAA (y) \equiv \frac{\int_{p_{T,\text{min}}}^{p_{T,\text{max}}} 
d\pT^2 \, \RAA(\pT,y) E_T^{-4}}{\int_{p_{T,\text{min}}}^{p_{T,\text{max}}}d\pT^2 E_T^{-4}}. \label{zetaraa}
\end{equation}
For the rapidity average we take flat distribution. In the above definitions the impact parameter, $b$, dependence is implicit. In order to average over the centrality we first convert impact parameter to centrality classes, $C$, using Glauber formalism. Then we integrate over centrality using probability function $e^{-C/20}$ which reproduces the experimental measurement \cite{Chatrchyan:2012np}. 
The procedure described above results in the so called ``raw'' $\RAA$ for each state. In order to calculate the inclusive $\RAA$ we have to take into account the feed-down from decays of the excited states. For that purpose construct the linear combinations $\RAA^{\Upsilon(1s)} =\sum_{i} f_i^{\Upsilon(1s)}$ and $\RAA^{\Upsilon(2s)} =\sum_{i} f_i^{\Upsilon(2s)} \RAA^{i, \text{raw}}$ where we use recent $\pT$-averaged feed-down fractions from ATLAS, CMS and LHCb data measured in p-p collisions. In particular we use
$f_i^{\Upsilon(1s)} = \{ 0.618,0.105,0.02,0.207,0.05 \}$ for
 $i\in \{ \Upsilon(1s) ,  \Upsilon(2s) ,  \Upsilon(3s) ,  \chi_{b1} ,   \chi_{b2} \}$, respectively. For
$f_i^{\Upsilon(2s)} = \{0.5,0.5\}$ we take
$i\in \{   \Upsilon(2s) ,  \Upsilon(3s)\}$ \cite{Strickland:2012as}.    

\begin{figure}[t]
\centering
\includegraphics[width=7cm,clip]{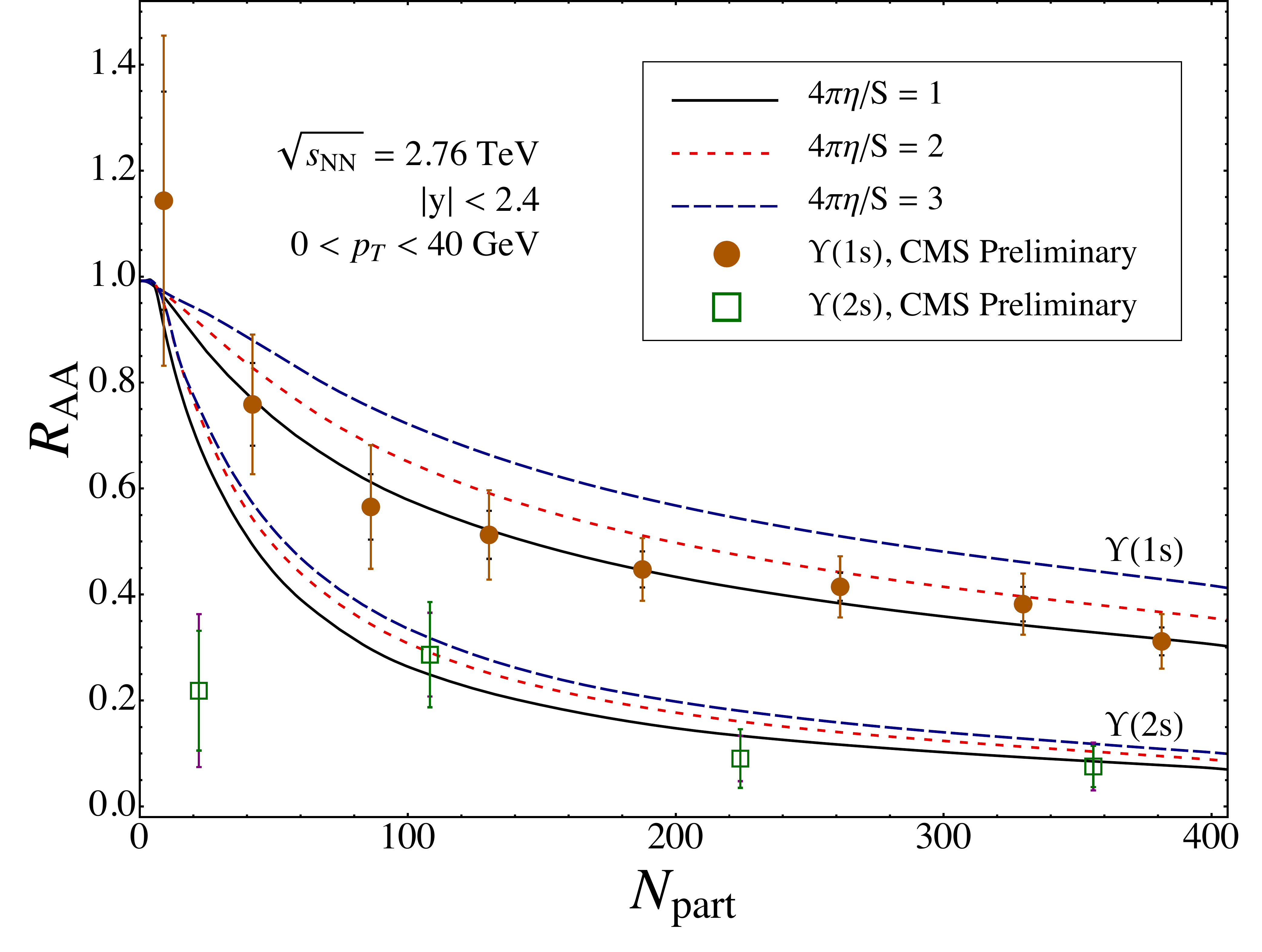}
\includegraphics[width=7cm,clip]{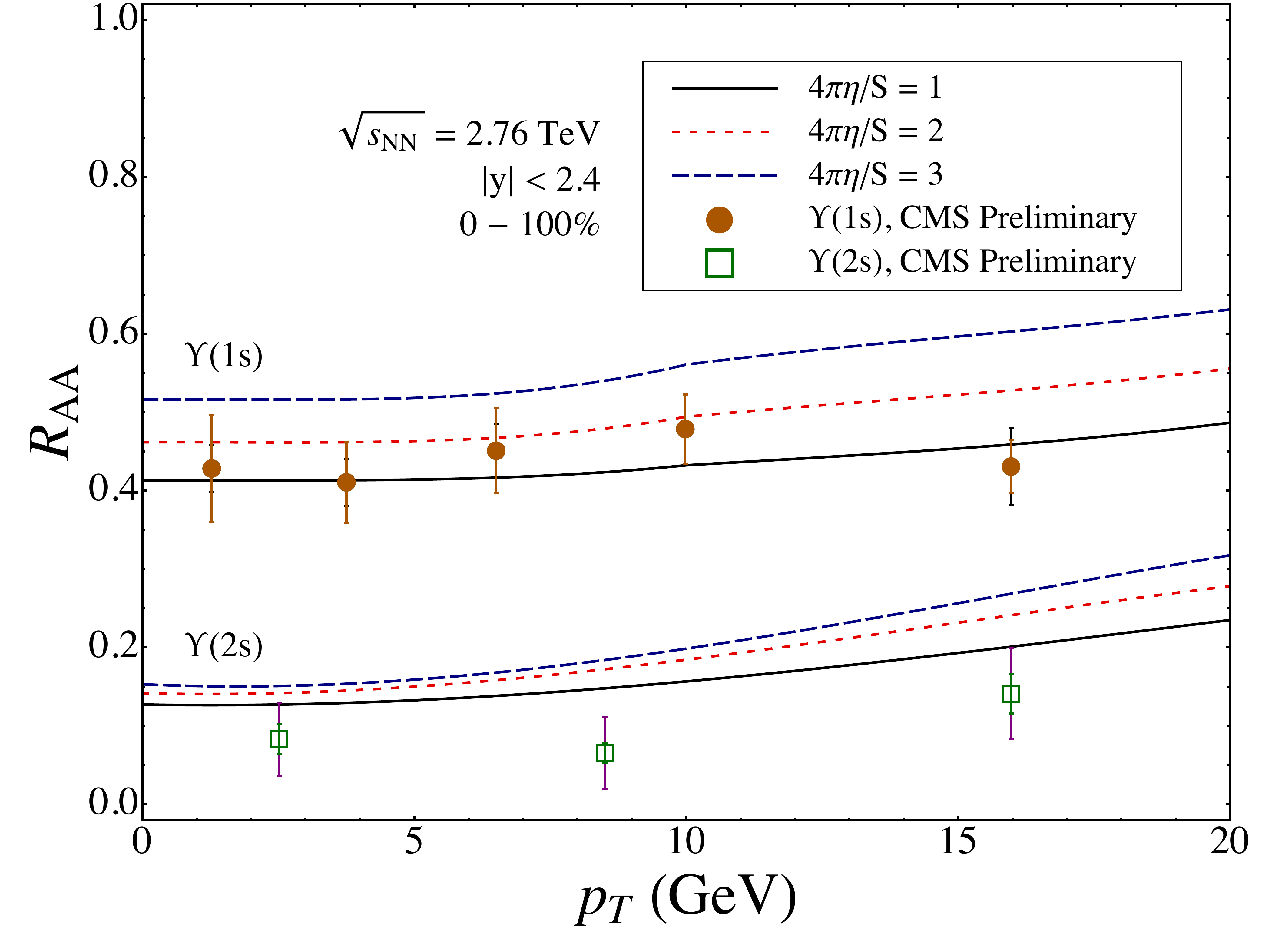}
\caption{(Color online) Inclusive $R_{AA}$ for the $\Upsilon(1s)$ and $\Upsilon(2s)$ states for different values of $\eta/s$ as a function of number of participants $N_{part}$ (left) and transverse momentum $\pT$ (right) compared to the CMS experimental data (symbols) \cite{CMS:2011ora}.}
\label{fig-2}       
\end{figure}

%
\section{Results}
\label{sec-5}
%
\par In the left panel of Fig.~\ref{fig-2} we present the comparison of the model predictions for the inclusive $\RAA$ factor of $\Upsilon(1s)$ and $\Upsilon(2s)$ states as a function of number of participants, $\Np$, with the preliminary experimental data from $2.76$ TeV Pb-Pb collisions measured by CMS at the LHC \cite{CMS:2011ora}. The results are obtained for various values of $\eta/s$. Our model seems to provide a reasonable description of the data except for $\Upsilon(2s)$  in the most peripheral collisions.
\par In the right panel of Fig.~\ref{fig-2} we present the analogue comparison as in the left panel except as a function of transverse momentum. We again observe an overall good agreement with the data, with the slow rise at large $\pT$ resulting from the time dilation of the formation times. Our result suggests that the underlying assumption that the states are decoupled from the QGP is in a good agreement with reality. Both, $N_{part}$ and $\pT$, dependence prefer small values of $1<4 \pi\eta/s < 2$, which stays in agreement with the recent hydrodynamic fits to light hadron correlation data.
%
\section{Conclusions}
\label{sec-conc} 
%
\par In this proceedings contribution we briefly presented our recent results on the thermal suppression of bottomonia in the anisotropic quark-gluon plasma produced in $2.76$ TeV Pb-Pb collisions at the LHC \cite{Krouppa:2015yoa}. For the study we used a pNRQCD model developed in Refs.~\cite{Strickland:2011mw,Strickland:2011aa} and upgraded in Refs.~\cite{Krouppa:2015yoa, Krouppa:2016jcl}  to include: (a) realistic (3+1)-dimensional QGP evolution within anisotropic hydrodynamics approach \cite{Ryblewski:2015hea}, (b) updated mixing fractions of the bottomonia states recently measured by ATLAS, CMS, and LHCb, and (c) improved centrality averaging procedure. The presented results on the $\RAA$ suppression factor show reasonable agreement with the data. Based on the comparison of number of participants $N_{part}$ and transverse-momentum $\pT$ dependence of the model inclusive $\RAA$ suppression factor with the recent CMS experimental data \cite{CMS:2011ora} we find that the values of $\eta/s$ lie in the range between $1/(4\pi)$ and $2/(4\pi)$. These values are in agreement with the most recent fluid dynamical fits to the light hadron correlation data, which  confirms that the QGP produced at the LHC energies is an almost-perfect fluid.
%
\begin{acknowledgement}
%
Author would like to thank Michael Strickland and Brandon Krouppa for the fruitful collaboration. This work was  supported by the Polish National Science Center grant No. DEC-2012/07/D/ST2/02125.
%
\end{acknowledgement}

\bibliography{refs}

\end{document}